# Inter-vendor harmonization of CT reconstruction kernels using unpaired image translation

Aravind R. Krishnan[a], Kaiwen Xu[b], Thomas Li[c], Chenyu Gao[a], Lucas W. Remedios[b], Praitayini Kanakaraj[b], Ho Hin Lee[b], Shunxing Bao[a], Kim L. Sandler[g], Fabien Maldonado[f,h], Ivana Išgum[i,j,k], Bennett A. Landman[a,b,c,d,e]

[a]Department of Electrical and Computer Engineering, Vanderbilt University, TN, USA, [b]Department of Computer Science, Vanderbilt University, Nashville, TN, USA, [c]Department of Biomedical Engineering, Vanderbilt University, Nashville, TN, USA, [d]Department of Radiology and Radiological Sciences, Vanderbilt University Medical Center, Nashville, TN, USA, [e]Vanderbilt University Institute of Imaging Science, Vanderbilt University Medical Center, Nashville, TN, USA, [f]Department of Medicine, Vanderbilt University Medical Center, Nashville, TN, USA, [g]Department of Radiology, Vanderbilt University Medical Center, Nashville, TN, USA, [h]Department of Thoracic Surgery, Vanderbilt University Medical Center, Nashville, TN, USA, [i]Department of Biomedical Engineering and Physics, Amsterdam University Medical Center, University of Amsterdam, Amsterdam, Netherlands, [j]Department of Radiology and Nuclear Medicine, Amsterdam University Medical Center, University of Amsterdam, Amsterdam, Netherlands, [k]Informatics Institute, University of Amsterdam, Amsterdam, Netherlands

## ABSTRACT

The reconstruction kernel in computed tomography (CT) generation determines the texture of the image. Consistency in reconstruction kernels is important as the underlying CT texture can impact measurements during quantitative image analysis. Harmonization (i.e., kernel conversion) minimizes differences in measurements due to inconsistent reconstruction kernels. Existing methods investigate harmonization of CT scans in single or multiple manufacturers. However, these methods require paired scans of hard and soft reconstruction kernels that are spatially and anatomically aligned. Additionally, a large number of models need to be trained across different kernel pairs within manufacturers. In this study, we adopt an unpaired image translation approach to investigate harmonization between and across reconstruction kernels from different manufacturers by constructing a multipath cycle generative adversarial network (GAN). We use hard and soft reconstruction kernels from the Siemens and GE vendors from the National Lung Screening Trial dataset. We use 50 scans from each reconstruction kernel and train a multipath cycle GAN. To evaluate the effect of harmonization on the reconstruction kernels, we harmonize 50 scans each from Siemens hard kernel, GE soft kernel and GE hard kernel to a reference Siemens soft kernel (B30f) and evaluate percent emphysema. We fit a linear model by considering the age, smoking status, sex and vendor and perform an analysis of variance (ANOVA) on the emphysema scores. Our approach minimizes differences in emphysema measurement and highlights the impact of age, sex, smoking status and vendor on emphysema quantification.

**Keywords:** Deep learning, image translation, generative adversarial networks, harmonization, computed tomography

## 1. INTRODUCTION

Image resolution and noise in computed tomography (CT) scans are dependent on raw data acquisition parameters and reconstruction parameters[1]. In the context of lung imaging using CT, the reconstruction kernel has an impact on emphysema quantification[2] and the robustness of radiomic features for different lung diseases[3]. There exists a tradeoff between spatial resolution and noise based on the choice of reconstruction kernel[4]. A hard kernel has higher spatial resolution accompanied by noise while a soft kernel has lower spatial resolution with reduced noise[4]. This trend is observed within a vendor and across vendors (**Figure 1**). The sharpness of the kernel affects the values of quantitative image features during image analysis, creating differences in measurements[5].

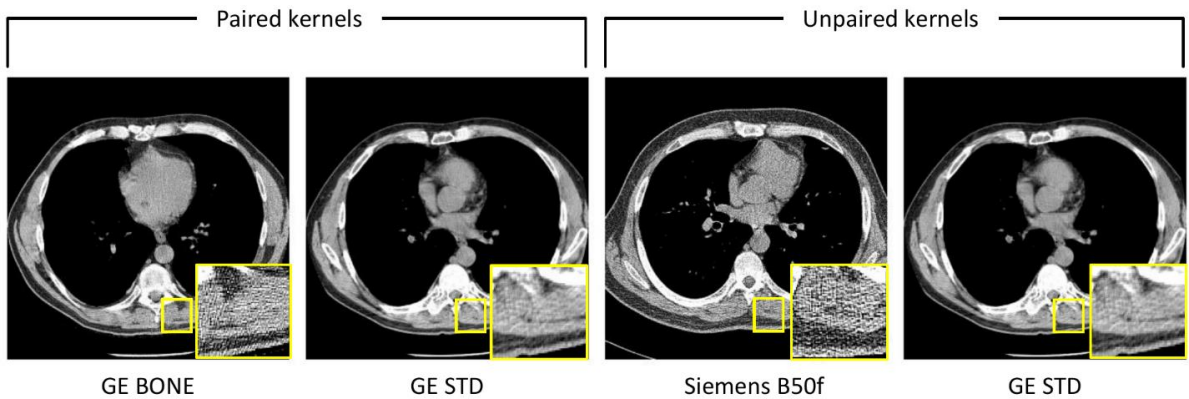

**Figure 1.** Differences in reconstruction kernels can be minimized by harmonizing to a reference standard. Harmonizing between paired kernels (left) has been explored due to the presence of one-to-one pixel correspondence between scans. However, unpaired kernels (right) create additional difficulties due to the difference in the anatomical alignment of scans obtained for different subjects from different vendors.

Kernel harmonization is a method that standardizes quantitative measurements across reconstruction kernels. Existing methods include physics-based and deep learning approaches. A physics-based harmonizer involving the modulation transfer function and global noise index was developed and its efficiency was evaluated on emphysema quantification[6]. Another physics-based approach implemented a generative deep learning model for harmonization and measured its performance on image similarity metrics and emphysema-based imaging biomarkers[7]. Juntunen et al.[8] investigated harmonization of image quality in computed tomography using reconstruction kernels and algorithms obtained from six different scanners and determined the noise power spectrum and modulation transfer function for the purpose of image harmonization. Deep learning approaches perform kernel conversion on paired data by learning the differences between high and low-resolution images using convolutional neural networks (CNN)[9]. Tanabe et al.[10] employed a harmonization method on paired hard and soft kernels using a CNN and evaluated the performance on emphysema, intramuscular adipose tissue and coronary artery calcification. Lee et al.[11] implemented kernel conversion from one kernel to various other kernels using CNNs. Fully convolutional networks (FCN) are also used for kernel harmonization. Bak et al.[12] implemented an FCN for image-to-image translation from a hard to soft kernel to study emphysema quantification. In addition to quantitative assessment, kernel conversion using CNN has shown to improve the reproducibility of radiomic features for pulmonary nodules and masses[13]. Dongyang et al.[14] developed a deep learning-based harmonization framework for kernel harmonization between reconstruction kernels obtained from patient and phantom data and showcased the importance of harmonization in radiomic feature reproducibility. These methods explore kernel harmonization in CT scans reconstructed with pairs of hard and soft kernels that have one-to-one mapping.

Generative adversarial networks (GANs) can generate synthetic images from Gaussian noise[15]. Image-to-image translation involves mapping a source image to a target image, preserving the contents of the source and transferring the style of the target[16]. Conditional GANs[17] have been employed for image translation in the form of pix2pix GAN for paired data[18] and cycle GAN for unpaired data[19]. Advanced generative models have been developed that can perform unsupervised image translation[20], multimodal image translation[21] and multidomain image translation[22]. In medical imaging, GANs have been used in clinical applications that include image synthesis, image reconstruction, cross-modality synthesis, image analysis and pseudo-healthy synthesis[23]. When considering CT scans from different vendors reconstructed with different kernels, a one-to-one mapping does not exist. One approach to harmonize across unpaired kernels is to use a cycle GAN where the goal is to translate images from the source to target domain and back to the source domain, ensuring a cycle consistent translation. Selim et al[24]. proposed a novel cycleGAN model for cross vendor harmonization between CT images from a Siemens scanner and GE scanner. The proposed model incorporated a convolution block attention map module in the generator followed by a domain loss between the synthesized and target domain images and was evaluated on standard radiomic features.

Using the concept of a cycle GAN, we implemented a multipath cycle GAN for kernel harmonization between kernels from the same vendor and kernels from different vendors. The generators were built using a combination of shared encoder-decoder architectures, creating multiple harmonization paths. This enables an encoder to share the latent space with the corresponding target decoders for multi domain image-to-image translation (**Figure 2**). We evaluated our model on emphysema quantification by standardizing hard and soft kernels to a reference soft kernel. We studied the effect of age, sex, smoking status and vendor on emphysema scores before and after harmonization by fitting a linear regression model and carrying out an analysis of variance (ANOVA) statistical test.

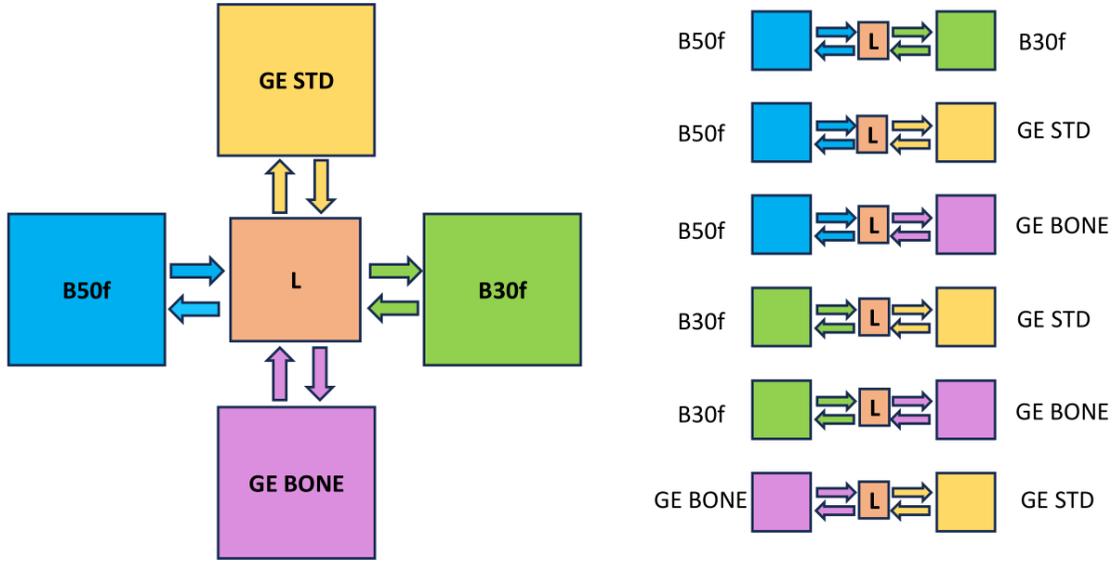

**Figure 2**. Kernel harmonization across four different reconstruction kernels can be performed using multiple cycle GANs operating across multiple paths. For a given source domain, a latent space is obtained from a source encoder that can be decoded by the corresponding target decoders. This approach enables harmonization between kernels from the same vendor and across kernels from different vendors using a high dimensional shared latent space (denoted as "L").

## 2. METHODOLOGY

We use data from the National Lung Screening Trial (NLST), a randomized controlled trial that compared low-dose CT (LDCT) scans of the chest with chest X-Ray in lung cancer screening[25]. Participants included in the trial were former and current smokers between the ages of 55 and 74 years, having a smoking history of at least 30 pack years[25]. We chose LDCT scans in the following manner: for every participant, the CT scans were reconstructed using different reconstruction kernels. Within a vendor, a participant had a scan reconstructed with a soft kernel and a hard kernel, forming a pair of scans. We consider the Siemens vendor consisting of the B50f (hard) kernel and B30f (soft) kernel and the GE vendor consisting of the BONE (hard) kernel and STD (soft) kernel. The peak kilovoltage output (kVp) for the scans from B50f, B30f, GE BONE and GE STD ranged from 80-140 kVp. We choose 50 scans for every reconstruction kernel resulting in a total of 200 scans to train the model. While testing our model, we consider 50 withheld scans each from the B50f, GE BONE and GE STD kernel and harmonize to the B30f kernel.

### 2.1 Pre-processing

We convert the CT scans from DICOM to NIfTI using the dcm2niix tool[26] (version 1.0.2). Before feeding the data to the model, the images are clipped to [-1024,3072] Hounsfield Units (HU) and normalized to [-1,1].

### 2.2 Multipath cycle GAN model

We use the concept of a cycle GAN and incorporate multiple paths to perform harmonization across different reconstruction kernels. To build a multipath kernel harmonization model, we use multiple generators and discriminators. The generator is a U-Net[27] with skip connections. We deconstruct the U-Net model into its respective encoder and decoder architectures, initializing encoders and decoders for every kernel. Skip connections enables the decoder to learn the low-level information (visual features) during the up-sampling process. A PatchGAN[18] is implemented as the discriminator model which classifies patches of images as real or synthetic. For a given path, the latent space from a source encoder is of size (512,1,1) where 512 represents the number of features obtained from the encoding process and the last two dimensions represent the spatial dimension. This latent space is utilized by three decoders from the respective target domains. In this fashion, each encoder shares its latent space with three other decoders depending on the path chosen for harmonization.

For the four different kernel domains, there are six possible directions to carry out harmonization: B50f to B30f, B50f to GE BONE, B50f to GE STD, B30f to GE STD, B30f to GE BONE and GE BONE to GE STD. A total of 12 different paths are created with six forward paths and six backward paths **(Figure 2)**. In every direction, we build the generator as follows: we treat one kernel as the source domain and use a source encoder to compress the image into a latent representation along with features from the down sampling process. We treat another kernel as the target domain and concatenate the features from the source encoder to the target decoder which decodes the latent representation, resulting in the generation of a synthetic image for the target domain (**Figure 3**). We stitch together all possible combinations of encoders and decoders, creating multiple generators for all paths. We use four different discriminators which are shared among all the reconstruction kernels depending on the direction of image translation.

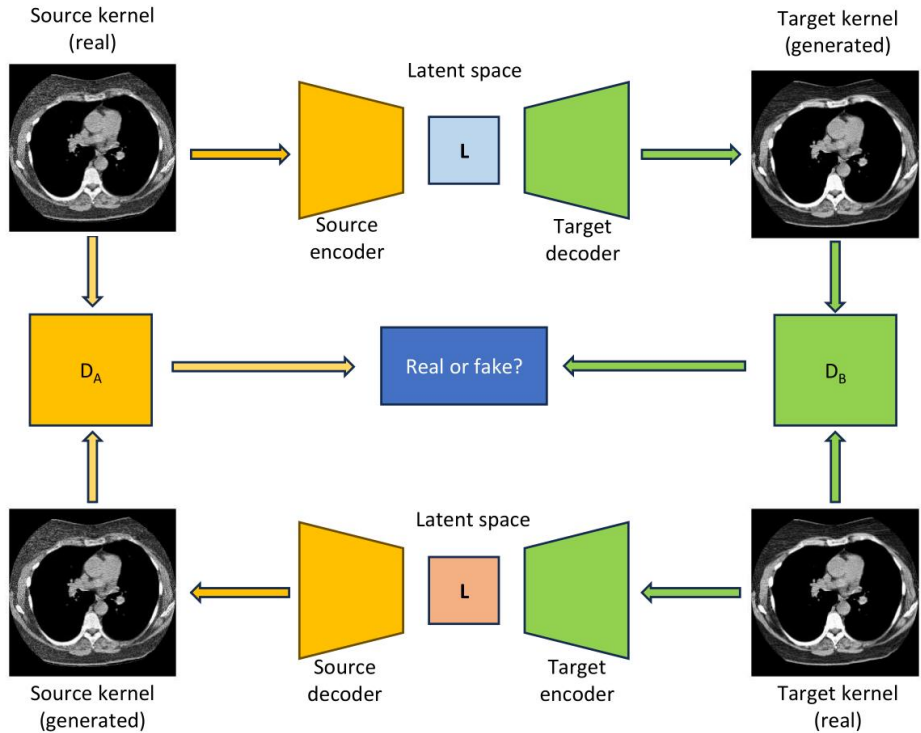

**Figure 3**. A cycle GAN consists of a forward and backward path. In the forward path, the source encoder and target decoder combine together to form a U-Net that generates a synthetic image with the style of the target domain. The synthetic image and the real target domain image are passed as inputs to a discriminator $D_B$ which distinguishes whether the generated image is real or fake. In the backward path, a synthetic image with the style of the source domain is generated which is fed to discriminator $D_A$ along with the source domain image.

Our model uses data from all the reconstruction kernels and is trained simultaneously on all paths. Our approach enables the model to harmonize from hard to soft kernels between and across vendors, hard to hard kernels and soft to soft kernels across vendors. We train our model on 2D grayscale axial slices of size 512 × 512 pixels from all domains. The images are loaded into the model at a size of 572 × 572 pixels and are cropped to 512 × 512 pixels. The model was trained in parallel on two Nvidia A6000 GPUs for a total of 30 epochs with a batch size of 8, with the Adam[28] optimizer and a learning rate of 0.0002. The learning rate remains constant for the first 15 epochs and begins to linearly decay for the next 15 epochs till it reaches 0. The generator and discriminator are governed by an adversarial loss which is implemented using the LSGAN[29] loss function. In addition to the adversarial loss, the generator is governed by an L1 cycle consistency loss. The weighting parameter, $\lambda$ for the adversarial loss is set to 10 for the forward and backward cycle paths. We use the default cycle GAN configuration of random horizontal flipping for data augmentation. A total of 12 adversarial losses, cycle losses and discriminator losses are implemented.

To validate our model, we estimate the ability of the model to minimize measurement differences in emphysema quantification. We compute lung masks for all the scans using an existing algorithm that automatically analysed the lung regions [30]. We compute percentage of voxels that have a radiodensity less than -950 HU using the segmented lung masks to obtain the emphysema score. We fit a linear model to estimate the effect of age, vendor, sex and smoking status on the emphysema scores for the respective kernels before and after harmonization. The linear regression equation is given by:

$$Y \sim \beta_0 + \beta_1 * X_1 + \beta_2 * X_2 + \beta_3 * X_3 + \beta_4 * X_4 + \varepsilon \quad (1)$$

where $Y$ is the emphysema measurement, $\beta_0$ is the intercept term, $X_1, X_2, X_3, X_4$ represent the age, sex, smoking status and vendor respectively, $\beta_1, \beta_2, \beta_3, \beta_4$ represent the regression coefficients of the independent variables and $\varepsilon$ is the error term.

## 3. RESULTS

We harmonize the B50f, GE STD and GE BONE kernels to the reference B30f kernel using the trained model. Prior to harmonization, each kernel has a different appearance due to the difference in textures that occur as a result of the vendor specific reconstruction. Harmonization standardizes the noise level to the reference kernel across all the reconstruction kernels (**Figure 4**). The converted B50f kernel, GE BONE kernel and GE STD kernel are translated using the style of the B30f kernel. The anatomy of the lung scans is preserved in both kernels after harmonization. Although the GE STD and GE BONE kernel are harmonized to the B30f kernel, artefacts are introduced in regions outside the lung field of view.

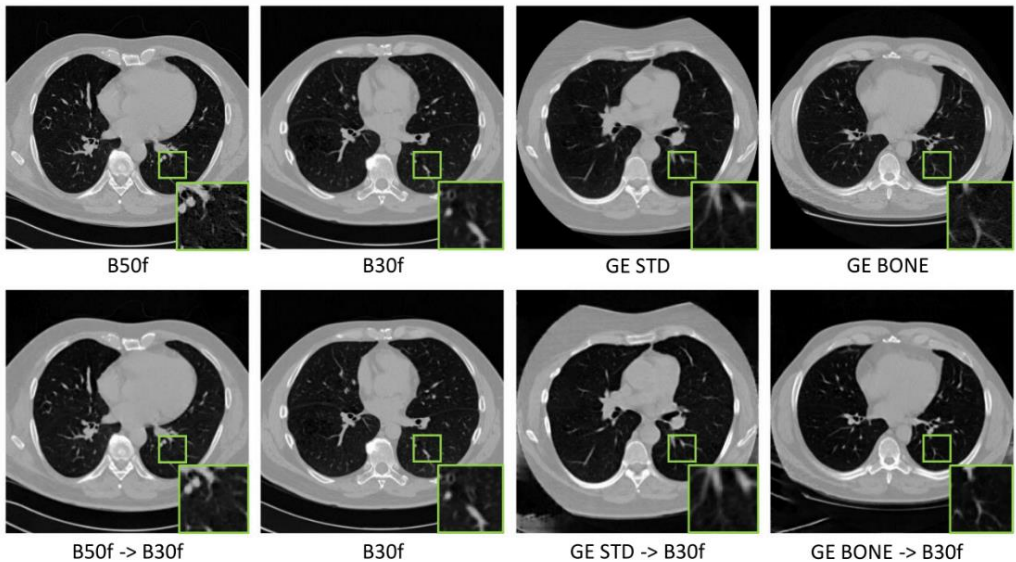

**Figure 4**. The noise in reconstruction kernels creates differences in the texture of underlying anatomical structures. The B50f and GE BONE hard kernels are noisy while the B30f and GE STD kernels are less noisy. Although the B30f and GE STD are soft kernels, their noise levels are different as these kernels belong to different vendors. Standardizing the GE soft kernel, GE BONE kernel and the B50f kernel to the reference B30f kernel (row 2) ensures consistent texture across all kernels for quantitative image analysis.

We assess the effect of harmonization on emphysema quantification. The emphysema scores are computed for subjects from different reconstruction kernels. Before harmonization, the range of emphysema scores of the B50f kernel, B30f kernel and the GE STD kernel are (2.16, 37.06), (0.27, 30.12) and (0.02, 20.02). After harmonization to the B30f kernel, the range of scores for the converted B50f kernel and the GE STD kernel are (0.15, 25.05) and (0.06, 25.4). The distribution of scores before and after harmonization are observed from violin plots (**Figure 5**). We also harmonized the GE BONE kernel to the B30f kernel but observed that the distribution of emphysema scores went from (1.02, 27.67) to (0.04, 43.84) (**Figure 5**). Although the GE BONE kernel resembles the B30f kernel in appearance, the emphysema scores are overestimated on the harmonized kernel as compared to the original emphysema score (**Figure 6**). For this reason, we exclude the GE BONE kernel from the regression analysis.

To study the effect of age, sex, vendor and smoking status on emphysema, we perform ANOVA after the models are fit to the data before and after harmonization. Age, sex and vendor had an impact on the emphysema scores for different reconstruction kernels before harmonization. Once the kernels were harmonized, vendor and sex are no longer significantly ($p>0.05$) related with the emphysema score while age continued to remain significantly ($p<0.05$) related. Smoking status had no impact before or after harmonization. All variables that are significantly related with emphysema score are highlighted in **Table 1**.

Table 1. ANOVA is performed on the emphysema scores before and after kernel harmonization. The effect of age, vendor, sex and smoking status are studied. All *p* values less than 0.05 are significant.

| **Parameters** | **Before harmonization** | **After harmonization** |
|---|---|---|
| Vendor | ***p* < 0.05** | *p* = 0.92 |
| Age | ***p* < 0.05** | ***p* < 0.05** |
| Sex | ***p* = 0.02** | *p* = 0.24 |
| Smoking status | *p* = 0.11 | *p* = 0.11 |

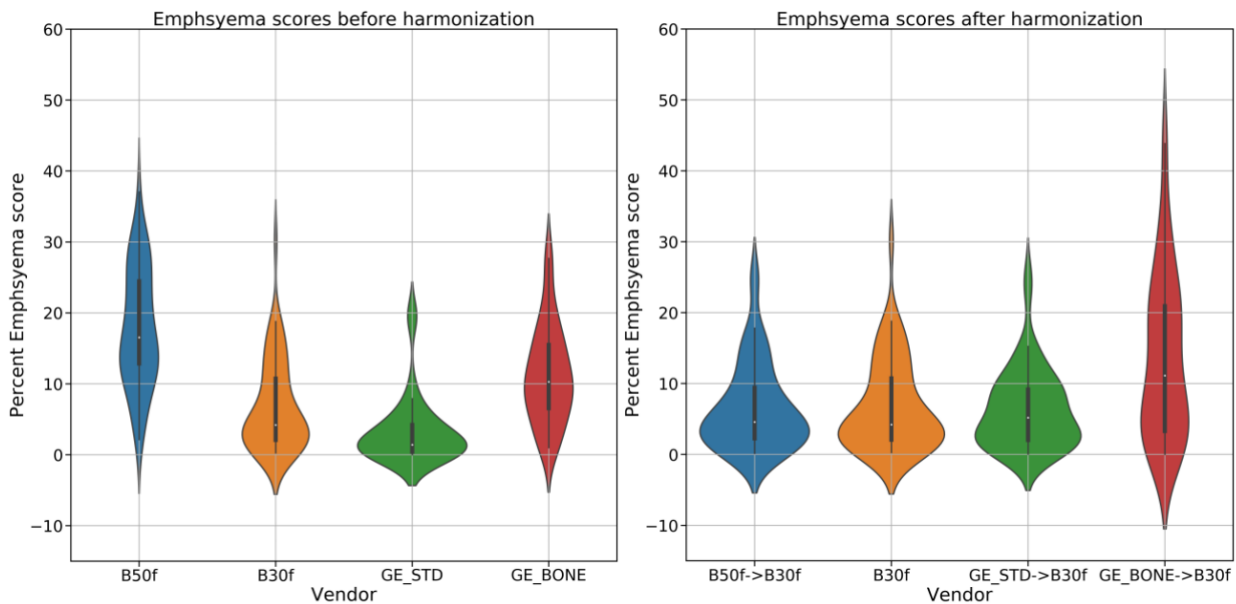

**Figure 5**. Percentage emphysema scores are affected by the reconstruction kernel in a given vendor, resulting in differences in measurements. Hard kernels overestimate emphysema quantification. Harmonizing kernels from different vendors to a reference soft kernel minimizes measurement errors, leading to a consensus among vendors for emphysema measurement.

# 4. DISCUSSION AND CONCLUSION

In this study, we investigated kernel harmonization in a multi-vendor, multi-kernel scenario by considering hard and soft reconstruction kernels from the Siemens and GE vendor. We implemented a multipath cycle GAN that can harmonize across different kernels in six different directions. We observe the efficiency of harmonization by standardizing the B50f (hard) kernel, GE BONE (hard) kernel and GE STD (soft) kernel to the B30f soft kernel and further evaluate emphysema quantification on the converted kernels. For Siemens, we observed that the model is able to convert the B50f kernel to the B30f kernel in an unpaired fashion on paired data. Across the vendors, the model was able to harmonize the GE STD kernel images by learning the style of the B30f. Prior to harmonization, there is variation in emphysema quantification between hard and soft kernels. This variation is minimized among most of the kernels after harmonizing to the reference B30f kernel as seen in **Figure 5**.

Our findings are consistent with previous studies that implemented kernel harmonization between paired kernels. Gallardo-Estrella et al.[31] showed that emphysema is sensitive to the reconstruction kernel. In their study, normalization of the kernel reduced the average differences in emphysema quantification for the Siemens and GE vendors. Additionally, Jin et al.[32] carried out a harmonization study on B50f and B30f kernels, showing that the lung density biomarkers for emphysema reduced considerably after harmonization. We also looked at the impact of age, vendor, smoking status and sex on emphysema quantification. Vendor had a high influence on emphysema prior to harmonization, suggesting variations in measurements as seen in **Table 1**. Once the kernels were harmonized, the influence of vendor and sex on emphysema was not significant, concluding that the difference in measurements were minimized across the kernels. Therefore, harmonizing kernels to a reference soft kernel mitigated the site effect on emphysema.

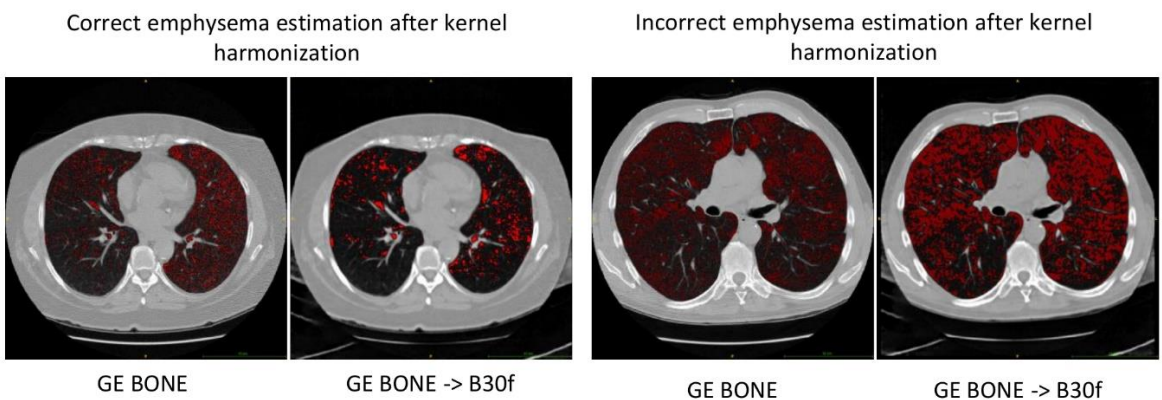

**Figure 6**. Emphysema quantification of GE BONE after kernel harmonization to B30f remained challenging. Although harmonization correctly reduced emphysema variation (left) in a few subjects, emphysema in the majority of subjects was over estimated (right).

Our approach has several limitations. We chose 50 scans for each reconstruction kernel while training the model. Although the model was trained on slices, the size of the dataset is limited, preventing the model from learning better representations. Furthermore, the harmonization of the GE BONE kernel was poor compared to other kernels while studying emphysema quantification. There were very few subjects where the emphysema variation was minimum. In one such case, the percent emphysema score reduced from 11.59% to 6.31%. In most of the cases, emphysema was over estimated after harmonization. A representative case for overestimation of emphysema can be seen in **Figure 6** where the emphysema score changed from 27.47% to 43.84%. Additionally, there were artefacts in the harmonized GE BONE and GE STD kernels that were observed outside the lung field of view. A possible explanation for this could be the lack of convergence of the model as a result of a small number of epochs for training. Additionally, the field of view (FOV) for the Siemens scans and GE scans are different. It is possible that the adversarial training could enforce the GE kernel images to represent the Siemens kernels image, resulting in artefacts outside the circular field of view. In future studies, better initialization, additional data for the current vendors, inclusion of additional vendors with different reconstruction kernels and longer epochs for training need to be implemented.

## ACKNOWLEDGEMENT

This research was funded by the National Cancer Institute (NCI) grant R01 CA253923. This work was also supported in part by the Integrated Training in Engineering and Diabetes grant number T32 DK101003. This research is also supported by the following awards: National Science Foundation CAREER 1452485; NCI grant U01 CA196405; grant UL1 RR024975-01 of the National Center for Research Resources and grant UL1 TR000445-06 of the National Center for Advancing Translational Sciences; Martineau Innovation Fund grant through the Vanderbilt-Ingram Cancer Center Thoracic Working Group; NCI Early Detection Research Network grant 2U01CA152662. The Vanderbilt Institute for Clinical and Translational Research (VICTR) is funded by the National Center for Advancing Translation Science Award (NCATS), Clinical Translational Science Award (CTSA) Program, Award Number 5UL1TR002243-03. The content is solely the responsibility of the authors and does not necessarily represent the official views of the NIH.